\documentstyle[12pt,epsf]{article}
\newcommand{\bc}{\begin{center}}
\newcommand{\ec}{\end{center}}
\newcommand{\bd}{\begin{displaymath}}
\newcommand{\ed}{\end{displaymath}}
\newcommand{\be}{\begin{equation}}
\newcommand{\ee}{\end{equation}}
\newcommand{\ba}{\begin{eqnarray}}
\newcommand{\ea}{\end{eqnarray}}
\newcommand{\bt}{\begin{tabular}}
\newcommand{\et}{\end{tabular}}

\begin{document}
\begin{flushright}
{\bf ITEP-PH--07-98}
\end{flushright}
\begin{flushright}
{\bf ISN 98--97}
\end{flushright}

\vspace{0.3cm}

\begin{center}
{\Large\bf
The $B_c$ Meson Lifetime in\\[5mm]
the Light--Front Constituent Quark Model }\\[1cm]
{\it A.Yu.Anisimov$^{(a)}$, I.M.Narodetskii$^{(a)}$,
C.Semay$^{(b)}$, and B.Silvestre--Brac$^{(c)}$}\\[1cm]
{\small $^{(a)}$Institute of Theoretical and Experimental Physics,
117259 Moscow,
Russia}\\[2mm]
{\small $^{(b)}$Universit\'{e} de Mons-Hainaut,  
Place du Parc, 20, B-7000 Mons, Belgium}\\[2mm]
{\small $^{(c)}$Institut des Sciences Nucl\'{e}aires, \\[2mm]
Avenue des Martyrs, 53, F-38026 Grenoble-Cedex, France}
\end{center}
\vspace{1cm}

\begin{abstract}
\noindent
We present an investigation of  the total decay rate of the (ground state) 
$B_c$ meson within the framework of the relativistic constituent quark model
formulated on the light-front (LF). The {\it exclusive} semileptonic (SL) and
nonleptonic (NL) beauty and charm decays of the $B_c$  meson are described 
through vector and axial hadronic form factors, which are calculated in terms 
of quark model LF wave functions. The latter ones are derived via the 
Hamiltonian LF formalism using as input the update constituent quark models. 
The {\it inclusive} SL and NL partial rates are calculated within a convolution 
approach inspired by the partonic model and involving the same $B_c$ wave 
function which is used for evaluation of the exclusive modes. The framework 
incorporates systematically 84 exclusive and 44 inclusive partial rates 
corresponding to the underlying $\bar{b}\to \bar{c}$ and $c\to s$ quark decays. 
We find $\tau_{B_c}=0.59\pm0.06$~ps where the theoretical uncertainty is
dominated by the uncertainty in the choice of LF wave functions and the 
threshold values for the hadron continuum. For the branching fractions of 
the $B^+_c \to J/\psi\mu^+\nu_{\mu}$ and $B_c^+\to J/\psi\pi^+$ decays we 
obtain $1.6 \%$ and $0.1\%$, respectively.
\end{abstract}

\vspace{2cm}


\newpage
\noindent The theoretical interest in the study of the $B_c$ meson, 
the bound state of the $\bar{b}c$ system with open charm and beauty, 
is stimulated by the experimental search at CDF and LHC. 
Recently the CDF Collaboration reported the observation of $B_c$ in 1.8
TeV 
$p\bar{p}$ collisions at Fermilab \cite{CDF98}. The CDF results for the 
$B_c$ mass and lifetime are $M_{B_c}=6.40\pm 0.39(stat) \pm
0.13(syst)~ \rm{GeV/c^2}$ and $\tau_{B_c}=0.46^{+0.18}_{-0.16}(stat)\pm 0.003
(syst)$ ps. The physics of $B_c$
mesons 
has stimulated much recent works on their properties, weak decays and 
production cross section on high energy colliders.
For a review see
\cite{GKLT95}. 

Similar to $D$ and $B$ mesons the ground $\bar{b}c$ state is stable
against strong or electromagnetic decay and
disintegrates only via weak interactions. The weak $B_c$ decays occur 
mainly through 
the CKM favored $\bar{b} \to \bar{c}W^+$ transitions with $c$ being a
spectator, leading to final states like $J/\psi\ell\nu$, and $c \to
sW^+$ 
transitions with $\bar{b}$ being a
spectator, leading to final states like $B_s\pi$, $B_s\ell\nu$. 
Weak decays of charmed and bottom hadrons are particularly simple in the
limit of infinite heavy quark mass, where the decay rate of a hadron
$H_Q$ containing a heavy quark $Q$ is completely determined by the decay rate
of the heavy quark itself. In this limit, one might expect that
$\Gamma(B_c) \approx \Gamma(\bar B^0)+\Gamma (D^0)$ yielding  
$\tau_{B_c} \approx 0.3$~ps, with $c$--decay dominating over
$\bar{b}$--decay \cite{B96}. 
In reality, hadrons are bound states of heavy quarks with light
constituents. 
The account of the soft degrees of freedom generates important
pre--asymptotic 
contributions due to the Fermi motion of a heavy quark inside the
hadron. These effects have a significant impact on the lifetime and 
various branching fractions of $B_c$. 
The various calculations of $\tau_{B_c}$ have been reported in the 
literature \cite{LM91}--\cite{BB96}. The wide range of predicted lifetimes 
$\tau_{B_c}=0.4-0.9$~ps, reflects the
uncertainty due to the various model
assumptions on the modification of the free decay rates due to the bound
state
effects and the limited knowledge of the heavy quark masses.

In this paper, we use an approach in which non--perturbative QCD effects 
are mocked up by a light--front (LF) 
wave function of the hadron \cite{MTM96}.  
The internal motion of a heavy
$Q$--quark inside the heavy flavour meson $H_Q$ is described by the 
distribution function $F(x)=\int d^2p_{\bot}|\psi(x,p^2_{\bot})|^2$, 
where $|\psi(x,p^2_{\bot})|^2$ represents the
probability to find a quark $Q$ carrying a LF fraction $x=p^+_Q/P^+_{H_Q}$ 
of the meson momentum and a transverse relative momentum squared $p_{\bot}^2$.
A relevant feature of this approach is that both exclusive and inclusive
decays are coherently treated in
terms of the same heavy quark wave function. 
So far the approach has been applied only for the exclusive and inclusive 
partial widths of $\bar{B}^0$ \cite{MTM96}, 
where it has been found 
that the overall picture is quantitatively satisfactory. 
Here we extend 
previous calculations started in Ref. \cite{AKNT98} to compute the lifetime 
and various decay branching fractions of $B_c^+$. Our aim is to constrain 
the model dependence of the calculated $\tau_{B_c}$ related 
to various choices of $\psi(x,p^2_{\bot})$. To this end we will make use of 
different LF wave functions, constructed via the Hamiltonian LF 
formalism (see e.g. 
\cite{C92}) adopting recently developed relativized and a non--relativistic 
quark models \cite{SS,SS98}. 

To start with, we briefly remind the main points of the procedure used to 
calculate heavy meson partial widths. For more details see 
Refs. \cite{MTM96}, \cite{AKNT98}. 
Consider the SL decay rates first. Instead of considering
the exclusive modes individually we will sum over all possible hadronic
final states $X$. This sum includes
hadronic states with a large range of invariant mass $M_X$. 
For the heavy--flavour mesons, the energy which flows into hadronic 
system is typically much larger than the energy scale $\Lambda_{QCD}$ which 
characterizes the strong interactions. Consequently, for a wide range of the 
phase space an inclusive description based on quark-hadron duality 
is appropriate. It will be valid over almost all of 
the Dalitz plot, failing only in the narrow corner region where the observed
mass spectrum is dominated by the two narrow peaks corresponding 
to the transitions $H_Q \to P$ and $H_Q \to V$, 
where $P$ and $V$ are the lowest lying pseudoscalar and vector mesons. 
Accordingly, the total SL rate of the $H_Q$ meson has been represented in the
following hybrid form
\be
\label{1}
\Gamma(H_Q \to X\ell\nu_{\ell})=
\Gamma(H_Q\to P\ell\nu_{\ell})+\Gamma(H_Q \to V\ell\nu_{\ell})+
\Gamma(H_Q\to X'\ell\nu_{\ell}),
\ee
where $X'$ represents the hadron continuum including also
the resonance states higher than $P$ and $V$.
The usefulness of such an
expansion rests on large {\it energy release} in the inclusive decay.

Our calculations of the {\it exclusive} rates 
$\Gamma(H_Q\to P\ell\nu_{\ell})$ and $\Gamma(H_Q \to V\ell\nu_{\ell})$ 
use the hadronic form factors 
that depends explicitly on dynamics of specific channels. 
The relevant formulae valid 
also in the case when the lepton masses are not negligible are
collected 
in \cite{AKNT98} and we do not quote them here. 
Instead, we will concentrate on the calculation of 
 the {\it inclusive} rate 
$\Gamma(H_Q \to X'\ell\nu_{\ell})$. 
The modulus squared of the amplitude summed over the final hadronic
states is written as
$|M|^2=(G^2_F/2)|V_{Q'Q}|^2L^{\alpha\beta}W_{\alpha\beta}$, 
where $V_{Q'Q}$ is the relevant CKM matrix element,
$L^{\alpha\beta}$ is the leptonic tensor
and $W_{\alpha\beta}$ is the hadronic tensor. We use the notation of Ref. 
\cite{BKSV94}. The hadronic tensor can be expressed in terms of  five structure
functions $W_1$ to $W_5$ that depend on two invariants, $q^2$ and
$q_0$, where $q$ is the 4--momentum of a lepton pair and $q_0$ 
is related to $M_X$ by: $q_0=(M_{H_Q}^2+q^2-M^2_X)/2M_{H_Q}$. 
It is convenient to scale all momenta by $M_{H_Q}$, letting $
M_X^2=s\cdot M_{H_Q}^2$ and $q^2=t\cdot M_{H_Q}^2$.
Evaluating the contraction $L^{\alpha\beta}W_{\alpha\beta}$ we arrive at 
the formula for the SL inclusive width
\be
\label{2}
\Gamma_{SL}=\frac{32}{3}\Gamma_0|V_{Q'Q}|^2J_{SL}\int\limits^{t_{max}}_{t_{min
}}
dt\Phi(t,m^2_1,m^2_2)\int\limits^{s_{max}}_{s_{min}}ds
\frac{|{\bf q}|}{M_{H_Q}}G(t,s),
\ee
where the prefactor $\Gamma_0=(G^2_FM^5_{H_Q})/(4\pi)^3$
sets the overall scale of the rate, and $J_{SL} \approx 0.9$ 
represents the effect of the radiative corrections \cite{NIR89}. 
In Eq. (\ref{2}) $\Phi(q^2,m^2_1,m^2_2)=
\sqrt{1-2\lambda_++\lambda^2_-}$, with $\lambda_{\pm}=(m^2_1\pm
m^2_2)/q^2$, $m_{1,2}$ being the lepton masses, 
$\lambda_1=\lambda_+-2\lambda^2_-$, $\lambda_2=\lambda_+-\lambda^2_-$. 
Furthermore, $2|{\bf q}|/M_{H_Q}\equiv \alpha(t,s)=\sqrt{(1+t-s)^2-4t}$,
and the limits of integrations in the $t-s$ plane are given by
$s_{min}=(M_X^{(0)}/M_{H_Q})^2,~~s_{max}=1-\sqrt{t},~~~~
t_{min}=(m_1+m_2)^2/m_Q^2,~~t_{max}=
(1-M_X^{(0)}/M_{H_Q})^2$, where $M_X^{(0)}$ is the threshold value at which 
the hadronic continuum starts.
The function $G(s,t)$ is expressed in terms of the linear combination of 
$W_i(t,s)$\footnote{The function $W_3$ does not appear in the total SL width 
integrated over the lepton 
energy \cite{BKSV94}.} :
\be
\label{3}
G(t,s) = 3t(1+\lambda_1-2\lambda_2)W_1+\left((1+\lambda_1)
\frac{{\bf
q}^2}{M_{H_Q}^2}+\frac{3}{2}\lambda_2t\right)W_2+
\frac{3}{2}\lambda_2t\left((1+t-s)W_4+tW_5\right).
\ee
To calculate the structure functions $W_i$ we use 
the parton approach
of Refs. \cite{MTM96} (see also \cite{JP}) based on the 
hypothesis of quark--hadron duality. This hypothesis assumes that 
the inclusive decay probability for which 
no reference to a particular hadronic state is needed 
equals to one into the free quarks. The basic ingredient 
is the expression for the hadronic tensor $W_{\alpha\beta}$ 
which is given through the optical theorem 
by the imaginary part of the quark box diagram describing the forward
scattering amplitude:
\be
\label{4}
W_{\alpha\beta}=\int\limits^1_0\frac{dx}{x}
w_{\alpha\beta}(p_{Q'},p_Q)\theta(\epsilon_{Q'})\delta[(p_Q-q)^2-m_{Q'}^2]
F(x),
\ee
where $w_{\alpha\beta}(p_{Q'},p_Q)=
\frac{1}{2}\sum\limits_{spins}\bar{u}_{Q'}O_{\alpha}u_b
\cdot \bar{u}_QO_{\beta}^+u_{Q'}$ is the parton matrix element squared. 
In what follows we choose the purely logitudinal kinematics with 
${\bf q}_{\bot}=0$. 
In Eq. (\ref{4}) the $\delta$--function corresponds to the decay of a
$Q$--quark with momentum $p_Q=xP_{H_Q}$ to a $Q'$--quark 
with the momentum $p_{Q'}=p_Q-q$ and
has two roots in $x$, {\it viz.} 
$\delta[(p_Q-q)^2-m^2_{Q'}]=[\delta(x-x_+)+\delta(x-x_-)]/
(M_{H_Q}^2|x_+-x_-|)$,
where $x_{\pm}(t,s)=\frac{1}{2}(1+t-s\pm\sqrt{(1+t-s)^2-4t
+4m^2_{Q'}/M^2_{H_Q}})$ and the quark transverse momenta being neglected
\footnote{By the quark masses $m_Q$ and $m_{Q'}$ we 
hereafter understand
the ``constituent" quark masses taken from a particular constituent quark
model.}.
The root $x_-$ is related to the contribution of
the $Z$--graph arising from the negative energy components of the
$Q'$--quark propagator and is prohibited by
the $\theta(\epsilon_{Q'})$ in Eq. (\ref{4}). Using the explicit expression 
for $w_{\alpha\beta}$ in Eq. (\ref{4}) one obtains
\be
\label{5}
W_1=F(x_+),~~
W_2=4\frac{x_+F(x_+)}{|x_+-x_-|},~~
W_3=W_4=-2\frac{F(x_+)}{|x_+-x_-|},~~
W_5=0.
\ee
In this way, the SL inclusive width gets related to the distribution $F(x)$. 
Substituting $W_i$ from (\ref{5}) into  (\ref{3}) we obtain
\ba
\label{6}
\Gamma(H_Q \to X'\ell\nu_{\ell})
&=&
\frac{2}{3}\Gamma_0J_{SL}
|V_{QQ'}|^2
\int\limits^{t_{max}}_{t_{min}}
dt\Phi(t)\int\limits^{s_{max}}_{s_{min}}ds\alpha(t,s)
\nonumber\\
&&
[(1+\lambda_1)\alpha^2(t,s)
\frac{x_+}{x_+-x_-}+3t(1-\lambda_-^2)]F(x_+).
\ea

The calculation of the NL decay rate closely follows the SL
one. We expect $H_Q$ decays to multimeson states to proceed
predominantly
via the formation of a quark--antiquark state, followed by the creation
of the additional $q\bar q$ pairs from the vacuum.
The effective weak Lagrangian, e.g.
for $\bar{b} \to \bar{c} u\bar{q}$ processes with
$q=d,s$ is given by
$L(\mu)=\frac{G_F}{\sqrt{2}}V_{cb}V_{uq}(c_1O_1+c_2O_2)$,
where $O_1$ and $O_2$ denote current--current operators with
the color--non--singlet and color--singlet structure, respectively. 
The lepton pair is substituted by a quark pair, and the Wilson
coefficients $c_i(\mu)$ are the perturbative QCD corrections
that describe the physics between the $W$ 
boson mass and the characteristic hadronic scale $\mu\approx m_Q$ 
of the process. 
We shall use the values \cite{BB93} 
$c_1(m_b)=1.132,~~c_2(m_b)=-0.286;~~c_1(m_c)=1.351,~~c_2(m_c)=-0.631$,
obtained at next--to--leading order with the evolution
of the running coupling constant being done at two--loop order using the
normalization $\alpha_s(m_Z)=0.118 \pm 0.003$. 

When calculating the NL decays we use again the hybrid approach. 
We write $\int dt$ in (\ref{2}) as 
$\int dt=\int dt \delta(t-\zeta_{P}^2)+
\int dt \delta(t-\zeta_{V}^2)+\int\limits_{t'_{min}}^{t_{max}} dt$, 
where $t'_{min} $ is now related to the threshold for the hadron continuum 
produced by the $W$ current, and $\zeta_{P,V}=M_{P,V}/M_{H_Q}$, with $M_{P,V}$ 
being the mass of a pseudoscalar or vector meson. 
The latter integral is treated  in the same way as for 
SL decays. According to Eq. (\ref{1}) it is the sum of 
{\it exclusive} rates and {\it inclusive} rates. 
The {\it inclusive} NL rate is given by Eq. (\ref{6})
with the substitution $\Gamma_0J_{SL} \to
3\Gamma_0|V_{uq}|^2\eta $ where 
$\eta=c^2_+(N_c+1)/(2N_c)+c^2_-(N_c-1)/(2N_c)$,
and $c_{\pm}=c_1\pm c_2$. In the limit of large number of colors $N_c$ 
which we use below $\eta=c^2_1+c^2_2$.

Additional contributions arise from the first two terms containing $\delta
(s-\zeta^2_{P,V})$. For the {\it exclusive} two--meson decays 
$H_Q\to PP,~PV,~VP$, and $VV$ 
we use the factorization approach \cite{BSW85} 
with the 
effective QCD coefficients $a_1=1,~a_2=-0.3$ and 
$a_1=c_1(m_c),~a_2=c_2(m_c)$ for the $\bar b\to \bar c$ 
and $c\to s$ transitions, respectively
\footnote{The coefficients $a_{1,2}$ correspond to two 
flavour--flow topologies relevant for our discussion: 
the so--called "tree topology" (class  I amplitudes), 
dominated by color allowed {\it external} $W$--decay
and the "colour--suppressed tree topology" (class II amplitudes), dominated by 
colour--supressed {\it internal} $W$--decays 
\cite{NS97}. There is the third class decays
in which the $a_1$ and $a_2$ amplitudes interfere. These decays play
the important role in the case of $D$ mesons, but are less important 
for $B_c$ and are absent for $\bar{B}^0$.}.       
The partial width of the {\it inclusive} NL decay $H_Q \to P+X'_{Q'Q}$ 
in which the final state
contains a charged or neutral {\it pseudoscalar} meson directly generated by a
color--singlet current are given by

\be
\label{7}
\Gamma_{PX}
=\frac{4\pi^2\eta f^2_P}{M_{H_Q}^2}\Gamma_0
|V_{q_1q_2}|^2|V_{Q_1Q_2}|^2
\cdot \int\limits^{(1-\zeta)^2}_{s_{min}}
ds [\tilde x_+(\tilde x_++\tilde x_-)^2-(\tilde x_-+3\tilde x_+)\zeta^2]
\frac{a(\zeta^2,s)F(\tilde x_+)}{|\tilde x_+-\tilde x_-|}.
\ee
The analogous inclusive width $\Gamma_{VX}$ for the production of 
a {\it vector} meson is obtained from (\ref{7}) 
by the substitution $f_P\to f_V$ 
and $[\tilde x_+(\tilde x_++\tilde x_-)^2-(\tilde x_-+3\tilde x_+)\zeta^2]
\to [\tilde x_+(\tilde x_++\tilde x_-)^2-(3\tilde x_-+\tilde x_+)\zeta^2]$.
Here $f_{P,V}$ are the pseudoscalar 
and vector meson coupling constants to the $W$ current, 
and $\tilde x_{\pm}=x_{\pm}(\zeta^2,s)$. 
The constants $f_{P,V}$ are taken from Ref. \cite{NS97}. 
For the $B_c$ meson mass we use $M_{B_c}=6.3$ GeV, the other meson masses 
are taken from the recent PDG publication \cite{PDG98}. 

The non--perturbative ingredient of our calculations is the LF wave
function
$\psi(x,p^2_{\bot})$. In what follows, we will adopt for the latter the
functions corresponding to the various equal time (ET) quark model wave
functions. There is a simple operational connection 
between ET and LF wave functions \cite{C92}, which allows to convert the
ET wave 
function $w(p^2)$ into relativistic LF wave function $\psi(x,p^2_{\bot})=
\frac{w(p^2)}{\sqrt{4\pi}}\frac{\partial p_z}{\partial x}$, 
where $p^2=p^2_{\bot}+p^2_z$, $p_z$ is the longitudinal momentum
defined as $p_z=(x-\frac{1}{2})M_0+\frac{m^2_{sp}-m^2_Q}{2M_0}$, 
and the free mass $M_0$ acquires the familiar form
$M_0=\sqrt{m_Q^2+p^2}+\sqrt{m_{sp}^2+p^2}$, with $m_{sp}$ being the 
mass of the quark--spectator. 

For the sake of brevity we shall present the results 
for the two representative ET wave functions corresponding 
to the AL1 \cite{SS} and DSR \cite{SS98} quark models.    
These functions result from solving either the Schr\"{o}dinger
equation (non-relativistic kinematics $T=p^2/2\mu_{red}$, $\mu_{red}=m_Qm_{sp}/
(m_Q+m_{sp})$; this is the case AL1)
or the Salpeter equation (semi-relativistic kinematics
$T=M_0$; this is the case DSR). The AL1 potential is the usual 
Coulomb+linear potential
supplemented with a gaussian hyperfine term whose range is mass
dependent. 
The DSR potential is much more sophisticated one since in addition to 
the previous 
terms, it contains also effects from instantons (which are necessary to 
describe light pseudoscalar mesons), and from finite quark size. 
The DSR potential is then convoluted with a quark density to
take into account finite size effects for quarks. The detailed form of
this potential is given in Ref. \cite{SS98}. 
The parameters of the AL1 and DSR potentials 
have been determined with help of a fit procedure on many experimental meson 
resonances, from $\pi$ to $B$. 

We choose these two different prescriptions
for the kinematics to see whether our results are sensitive to some
``relativization" at the level of the ET wave function. The main difference 
between $w(p^2)$ in the AL1 and DSR models 
relies in the behaviour at high value of the internal momentum. The AL1 model 
results in the soft wave function similar to that of the ISGW2 model 
\cite{IS95} with 
$<p^2>=0.235~{\rm GeV^2}$ for $\bar{B}^0$  
and
$<p^2>=1.075~{\rm GeV^2}$ for $B_c$, while the DSR wave function 
exibits high momentum components leading to 
$<p^2>=0.517~{\rm GeV^2}$ for $\bar{B}^0$  
and
$<p^2>=1.388~{\rm GeV^2}$ for $B_c$.  
As a result the DSR distribution function is broader than 
that for the case AL1. 
This is illustrated in Fig.1 which shows the distribution functions 
$F^{\bar b}(x)$ for $\bar B^0$ and $B_c$ mesons. 
Note that $F^c_{B_c}(x)=F^{\bar b}_{B_c}(1-x)$. 
In terms of the mean value $\bar{x}=\int\limits^1_0xF(x)dx$ and the
variance $\sigma^2=\int\limits^1_0(x-\bar{x})^2F(x)dx$, 
one obtains $\bar{x}_{B^0}=$ 0.9~(AL1), 0.89~(DSR), 
$\sigma^2_{B^0}=$ 0.0028~(AL1), 0.0065~(DSR),
$\bar{x}_{B_c}=$ 0.72~(AL1 and DSR), and $\sigma^2_{B_c}=$ 0.0062~(AL1), 
0.0081~(DSR). The constituent quark masses in a relativized quark models are 
systematically lower that those determined using the non--relativistic models. 
In our case $m_u=m_d=0.315,~m_s=0.577,~m_c=1.836,~m_b=5.227$ GeV (case AL1) 
and $m_u=m_d=0.221, m_s=0.434, m_c=1.686, m_b=5.074$ GeV (case DSR). 
Note that the {\it constituent} quark masses $m_b$ and $m_c$ satisfy
approximately the relation $m_b=m_c+3.4$~GeV, which is consistent with
the well known formula relating the {\it pole} masses $m_{b,pole}$ and
$m_{c,pole}$ in Heavy Quark Effective Theory.

We are now ready to present an overview over different $B_c$ decays 
and their relative importance as obtained within the framework we are
advocating. To this end we first calculate the partial $\bar{B}_0$ decay
modes corresponding to various underlying quark subprocesses. 
In Table 1 we report the $\bar{B}^0$ partial widths for AL1 and DSR LF models. 
For comparison, we show also the results obtained in  \cite{AKNT98} 
using the Gaussian--like ans\"atz of the ISGW2  model which are very similar 
to the AL1 case.
 
Our analysis incorporates 54 exclusive SL and NL decays
and 29 inclusive decays including two baryon-antibaryon channels.
The latter ones were calculated using the Stech approach \cite{S87}. 
We have also included the CKM suppressed $b\to u$ contributions 
with $|V_{ub}/V_{cb}|\approx 0.1$. 
The vector and axial form factors for $\bar{B}^0 \to
D(D^*)\ell\nu_{\ell}$ and $\bar{B}^0 \to \pi(\rho)\ell\nu_{\ell}$
transitions have been calculated using the formalism of Refs. \cite{DKND97}.
When calculating the {\it inclusive} rates the important question is 
which value to use for the hadron threshold 
$M^{(0)}_X$. 
This value is very badly defined theoretically. In our calculations we
have adopted two ``natural" choices: $M_X^{(0)}=M_{P}+m_{\pi}$ and
$M_X^{(0)}=M_V+m_{\pi}$. 
In Table 1 we show only the results obtained using 
$M_X^{(0)}=M_V+m_{\pi}$. For $M_X^{(0)}=M_P+m_{\pi}$ we obtain $\approx 10\%$ 
increasing of the calculated $\bar B^0$ rate. 
For each case we then fix the effective value of $|V_{cb}|
$ by the requirement that the measured $\bar B^0$ meson lifetime
$\tau_{\bar B^0} \approx 1.56$~ps is obtained.
Our goal here is not to
establish a new value of $|V_{cb}|$, but rather to illustrate how our
approach works. Moreover, imposing
the $|V_{cb}|$ constraint strongly reduces the dependence of the
predicted value of $\tau_{B_c}$ from the uncertainty
related to the choice of the continuous $c\bar{c}$ threshold \footnote
{The values of other CKM parameters entering calculations are 
$|V_{sc}|=0.974,~|V_{dc}|=|V_{us}|=0.221$.}.

We apply the same strategy to calculate different partial rates and the 
lifetime of the $B_c$.  The
critical point with regards to this issue is {\it inclusive} charm decay. 
Here the energy release is not as comfortably large as it is in the case of 
bottom decay. As a result, our estimations of the inclusive charm decay should 
be more sensitive to a hadronization model. However, this decay contributes 
only $\approx 10\%$ to the total $c\to s$ rate. For this reason we did not 
include any hadronization corrections in our calculations.

The results for the partial $B_c$ decay modes corresponding to the
various underlying quark subprocesses are collected in Table 2 for 
$M^{(0)}_X=M_V+m_{\pi}$. We also include non--spectator 
contributions from weak annihilation (WA) and Pauli interference (PI). 
The contribution of the annihilation channel is
\be
\label{ann}
\Gamma_{a}=\sum_{i=\tau,c}\frac{G^2_F}{8\pi}|V_{cb}|^2M_{B_c}^5
\left(\frac{f_{B_c}}{M_{B_c}}\right)^2\left(\frac{m_i}{M_{B_c}}\right)^2
\left (1-(\frac{m_i^2}{M_{B_c}})^2\right)^2\cdot \tilde c_i,
\ee
where we take $f_{B_c}\approx 0.5~$ GeV, 
$\tilde c_{\tau}=1$ for the $\tau^+\nu_{\tau}$
channel and $\tilde c=(2c_+(2\tilde{\mu}_{red})+c_-(2\tilde{\mu}_{red})^2/3$ 
for the $c\bar s$ channel, with $\tilde{\mu}_{red}=m_bm_c/(m_b+m_c)$ being the
reduced mass of the $\bar b c$ system. We use $c_+(\tilde{\mu}_{red})=0.8$, 
$c_-(\tilde{\mu}_{red})=1.5.$ For the $\Gamma_{PI}$ we use the expression given 
in \cite{VS87}.
Note that because of the substantial cancellation of weak
annihilation rate and the effect of the Pauli interference diagrams,
no significant uncertainty on the lifetime arises from the limited
knowledge of the decay constant $f_{B_c}$.
We find $\Gamma_a=0.268$~{\rm ps}$^{-1}$, $\Gamma_{PI}= -0.142$ ps$^{-1}$. 

We also show in Table 2 the results 
obtained in \cite{AKNT98} for the ISGW2 LF model \footnote{The result of 
Ref. \cite{AKNT98} corresponds to $f_{B_c}=0.42~{\rm{GeV}}$. After correcting 
for the value of $f_{B_c}$ used in this paper, one gets the result quoted 
in Table 2.} and in \cite{BB96},
using Operator Product Expansion (OPE). Viewing the latter comparison with due 
caution, regarding the model dependence and other uncertainties in the 
estimation of the decay modes as well as 
the quark mass uncertainty for the inclusive prediction, it is reassuring 
that the order of magnitude comes out to be consistent. 
Our bound state 
corrections are numerically larger than very small effects found in 
\cite{BB96}, especially for the $c\to s$ transitions.  One possible reason 
being is that an analysis of $c\to{s}$ decays using OPE is of 
limited validity, since due to smaller energy release the convergence of 
OPE is slower than for $\bar{b}\to\bar{c}$ decays.

For the sum of $\bar b \to \bar c$ spectator contributions we obtain 
$\Gamma^{(\bar b \to \bar c)}=0.501$~ps$^{-1}$ (AL1) and
$0.557$~ps$^{-1}$ 
(DSR), while the total $c$--decay contribution is $\Gamma^{(c\to s)}=
0.900$~ps$^{-1}$ (AL1) and $1.079$~ps$^{-1}$ (DSR).
One observes a dominance of the charm decay modes over $b$--quark
decays. Various branching
fractions can be also inferred from Table 2. For instance the
semileptonic
branching ratio $BR(B_c \to \ell\nu_{\ell} X)$ is found to be $9.3\%$. 
The absolute rates are 
$\Gamma(B_c\to \ell\nu_{\ell}X_{c\bar c})=4.0~(4.4)\cdot 
10^{13}|V_{cb}|^2~\rm {sec}^{-1}$ and 
$\Gamma(B_c\to \ell\nu_{\ell}X_{sb})=8.0~(9.5)\cdot 
10^{13}|V_{sc}|^2~\rm {sec}^{-1}$ for AL1 (DSR) cases, respectively.

Putting everything together one finds 
$\Gamma(B_c)=\Gamma^{\bar b \to \bar c}(B_c)+
\Gamma^{c \to s}(B_c)+\Gamma_a+\Gamma_{PI}=(0.65~{\rm ps})^{-1}$ (AL1),
and $(0.57~{\rm ps})^{-1}$ (DSR). For $M_X^{(0)}=M_P+m_{\pi}$ 
we obtain $\Gamma_{B_c}=(0.61~{\rm ps})^{-1}$ (AL1) and 
$(0.53~{\rm ps})^{-1}$ (DSR). We consider the 
dispersion in predicted values of $\tau_{B_c}$ as a rough measure of our 
theoretical uncertainty in calculation of inclusive decay rates arising  
both from the model dependence of $\psi(x,p^2_{\bot})$ and the different 
choices of the continuum threshold $M_X^{(0)}$.
Averaging our predictions we obtain
\be
\tau_{B_c}=(0.59\pm 0.06)~{\rm ps}.
\ee
This result compares 
favourably with estimates obtained using OPE \cite{BB96} and also agrees 
with the most recent CDF measurement within one standard deviation.

Finally, we note that the experimental extraction of $B_c$ signal in the
hadronic background
requires the reliable estimation of the branching fraction $
BR(B_c\to J/\psi+X) $, because $J/\psi$ can be easily identified by
its $\mu^+\mu^-$ decay mode, while the experimental registration of the
final states containing the $\eta_c$ or $B^{(*)}_s$ is impeded by the
large hadron background.  We obtain $ BR(B_c\to J/\psi+\pi)=0.1\%$ (AL1 
and DSR), and $BR(B_c\to J/\psi+X)=14.8\% $ (AL1), $13.6 \%$ (DSR). 
For the exclusive 
$B_c \to J/\psi\mu^+\nu_{\mu}$ channel whose signature would be quite clean 
experimentally when $J/\psi$ decays into a pair of muons, providing a three muon 
vertex the calculated branching fractions are $1.68\% $ and $1.57\%$ 
for cases AL1 and DSR, respectively.

In conclusion, adopting a LF constituent quark model we have investigated
the partial widths and the lifetime of the $B_c$ meson. The hadronic form
factors
and the distribution function were calculated from meson wave functions
derived from an effective $q\bar q$ interaction intended originally to
describe the meson mass spectra. In this way the link between $B_c$
physics and the ``spectroscopic" quark models was explicitly
established. For numerical estimates we have employed the LF functions 
related to the ET eigenfunctions of the AL1 and DSR quark models. In
several important aspects our analysis goes beyond the quark model
estimations obtained previously. In addition to the SL and the two--meson
NL {\it exclusive} decay modes  
we have included a number of the $B_c\to H_{\bar cc}(H_{\bar b
s})q_1q_2$ {\it exclusive}  and $B_c \to X'_{\bar cc}(X'_{\bar b
s})q_1q_2$ {\it inclusive} channels. These channels have considerable impact 
on the predicted overall $\bar b\to \bar c$ rate. 
To sum up, the LF constituent quark model makes clear predictions on the
global pattern: (i) a short $B_c$ lifetime $\approx 0.6$~ps and (ii) a
predominance of charm over beauty decays. 

\section*{Acknowledgements}
We are grateful for P.Yu.Kulikov for having checked the numerical results. 
Two of the authors (A.Yu.A. and I.M.N.) acknowledge the support of the 
INTAS--RFBR grants Refs. No 95--1300 and No 96--165. This work was done in 
part under the RFBR grant Ref. No 95-02-0408a.


\vspace{1cm}
\noindent {\bf Table 1 }. 
The partial widths of 
the $\bar{B}^0$ meson (in the units $|V_{cb}/0.039|^2{\rm ps}^{-1}$) 
calculated adopting the different choices of the LF 
wave functions. Each SL width is the sum of 2 {\it exclusive} and one 
{\it inclusive} widths. For a NL decay the width is the sum of 12 
{\it exclusive} and 6 {\it inclusive} channels corresponding to the 
external and internal $W$ decays. The CKM matrix element $|V_{cb}|$ is calculated 
using the experimental value of $\Gamma_{exp}(B^0)=0.641~{\rm ps}^{-1}$. 
Also shown are the SL branching ratios ${\rm BR_{SL}}$ and the charm counting 
$n_c$.\\[5mm]

{\Large
\begin{center}

\begin{tabular}{|c|c|c|c|}
\hline\hline
Model & AL1 & DSR & ISGW2 \cite{AKNT98}\\
\hline\hline
$\bar{b}\to \bar{c}+e\nu_e$ & 0.076 & 0.072 & 0.074\\
\hline
$\bar{b}\to \bar{c}+\mu\nu_{\mu}$ & 0.075 & 0.071 & 0.074\\
\hline
$\bar{b}\to \bar{c}+\tau\nu_{\tau}$ & 0.016 & 0.016 & 0.015\\
\hline
$\bar{b}\to \bar{c}+u\bar{d}$ & 0.322 & 0.311 & 0.313\\
\hline
$\bar{b}\to \bar{c}+c\bar{s}$ & 0.122 & 0.137 & 0.123\\
\hline
$\bar{b}\to \bar{c}+u\bar{s}$ & 0.016 & 0.016 & 0.016\\
\hline
$\bar{b}\to \bar{c}+c\bar{d}$ & 0.006 & 0.007 & 0.006\\
\hline
$\bar{b}\to \bar{u}$ & 0.007 & 0.007 & 0.007\\
\hline
$\bar{B}^0 \to N\bar{\Lambda}_c$, $\Lambda_c\bar{\Xi}_c$& 0.023 & 0.031 & 0.023\\
\hline
$\Gamma(\bar{B}^0)$ & 0.662 & 0.668 & 0.652\\
\hline
$|V_{bc}|$ & 0.0383 & 0.0382 & 0.0386\\
\hline
\hline
$n_c$ & 1.193 & 1.205 & 1.198\\
\hline
$BR_{SL}$ & $11.42\%$ & $10.78\%$ & $11.35\%$\\
\hline\hline
\end{tabular}
\end{center}
}
\newpage
\noindent {\bf Table 2 }. Inclusive partial rates of $B_c$ 
(in units ps$^{-1}$) for the choice of the continuum threshold 
$M_X^{(0)}=M_V+m_{\pi}$. The $b \to c$ rates are calculated using 
the effective value of $|V_{cb}|$ from Table 1. The $c \to s$ 
rates are calculated using $|V_{cs}|=0.974$. $\Gamma_a$ and $\Gamma_{PI}$ 
are calculated using $f_{B_c}=0.5~{\rm GeV}$.

{\Large\begin{center}

 {\bf Table 2 } \\[1.5cm]
\begin{tabular}{|c|c|c|c|c|}
\hline\hline
Model & AL1 & DSR & ISGW2 \cite{AKNT98} & OPE \cite{BB96}\\
\hline\hline
$\bar{b}\to \bar{c}+e\nu_e$ & 0.058 & 0.060 & 0.061 &0.075\\
\hline
$\bar{b}\to \bar{c}+\mu\nu_{\mu}$ & 0.058 & 0.060 & 0.061 &0.075\\
\hline
$\bar{b}\to \bar{c}+\tau\nu_{\tau}$ & 0.013 & 0.014 & 0.013 &0.018\\
\hline
$\bar{b}\to \bar{c}+u\bar{d}$ & 0.244 & 0.261 & 0.259 &0.310\\
\hline
$\bar{b}\to \bar{c}+c\bar{s}$ & 0.093 & 0.117 & 0.102& 0.137\\
\hline
$\bar{b}\to \bar{c}+u\bar{s}$ & 0.012 & 0.013 & 0.013 & \\
\hline
$\bar{b}\to \bar{c}+c\bar{d}$ & 0.005 & 0.006 & 0.006& \\
\hline
$\bar{b}\to \bar{u}$ & 0.005 & 0.007 & 0.005 & \\
\hline
$B_c\to \Sigma_c\bar\Sigma_c$ & 0.013 & 0.019 & 0.011 & \\
\hline
$\Gamma^{\bar{b}}$ & 0.501 & 0.557 & 0.531 &0.615\\
\hline\hline
$c\to s+e\nu_e$ & 0.078 & 0.090 & 0.081 & 0.162\\
\hline
$c\to s+\mu\nu_{\mu}$ & 0.073 & 0.085 & 0.077 & 0.162\\
\hline\hline
$c\to s+u\bar{d}$ & 0.688 & 0.814 & 0.698 & 0.905\\
\hline\hline
$c\to s+u\bar{s}$ & 0.024 & 0.029 & 0.023 &  \\
\hline
$c\to d$ & 0.037 & 0.061 & 0.036 & \\
\hline\hline
$\Gamma^{c}$ & 0.900 & 1.079 & 0.915 & 1.229\\
\hline\hline
$\bar{b}c\to \tau\nu_{\tau}$ &0.074 &0.074 & 0.074 &0.056\\
\hline
$\bar{b}c\to c\bar{s}$ &0.194 &0.194 & 0.194 & 0.138\\
\hline
PI &-0.142 &-0.142& -0.142 &-0.124\\
\hline\hline
$\Gamma_{tot}$ & 1.527 & 1.762 & 1.571 & 1.914\\
\hline
$\tau_{B_c}$ & 0.65 & 0.57 & 0.64 & 0.52\\
\hline\hline
\end{tabular}
\end{center}
}

\newpage
\vspace{0.5cm}
\begin{figure}[htb]
\centerline{\epsfxsize=7.0in \epsfysize=8.0in \epsffile{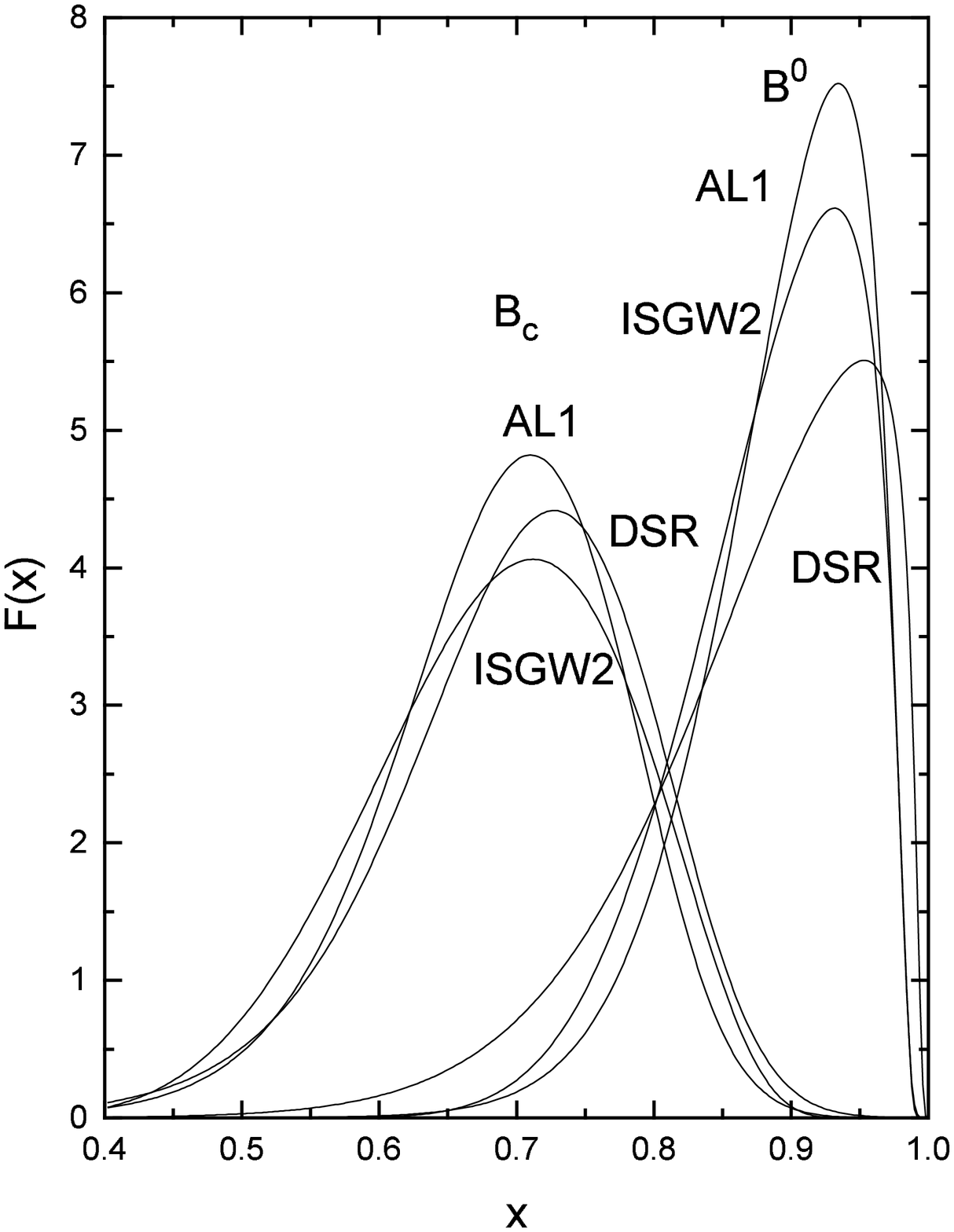}}
\caption {
The distribution functions $F_{\bar b}(x)=\int d^2
p_{\bot}|\psi_{\bar b}(x,p^2_{\bot})|^2$ for the $\bar 
{B}^0$ and $B_c$ mesons calculated using the AL1 and DSR LF wave functions 
(this work) and the ISGW2 LF wave function (Ref. \cite{AKNT98}).}
\end{figure}


\begin{thebibliography}{99}
\bibitem{CDF98} F.Abe, CDF Collaboration, Phys. Rev. Lett.  {\bf 81} 
(1998) 2432, hep--ex/9804014.
\bibitem{GKLT95} S.S.Gerstein, V.V.Kiselev, A.K.Likhoded, and
A.V.Tkabadze,
Usp. Fiz. Nauk {\bf 165} (1995) 3 [Phys. Usp. {\bf 38} (1995) 1],
hep--ph/9504319; Phys. Rev. D {\bf 51} (1995) 3613.
\bibitem{LM91} M.Lusignoli, M.Mazsetti, Z. Phys. C {\bf 51} (1989) 549; 
S.S.Gerstein {\it et al.}, Int. J. Mod. Phys. {\bf A6}
(1991) 2309; P.Cotangelo {\it et al.}, Z. Phys. {\bf C57} (1993) 43.
\bibitem{B96} I.Bigi, Phys. Lett. {\bf B371} (1996) 105.
\bibitem{BB96} M.Beneke and G. Buchalla, Phys. Rev. {\bf D53} (1996)
4991.
\bibitem{MTM96} V.L.Morgunov, K.A.Ter--Martirosyan, Phys. Atom. Nucl.
{\bf 59} (1996) 1221; I.L.Grach, I.M.Narodetskii, S.Simula,
K.A.Ter--Martirosyan,
Nucl. Phys. {\bf B502} (1997) 227; Nucl. Phys. B (Proc. Suppl.) {\bf 55A}
(1997) 84.
\bibitem{AKNT98} A.Anisimov, P.Yu.Kulikov, I.M.Narodetskii, and K.A. 
Ter--Martirosyan, hep--ph/9809249, to appear in Phys. Atom. Nucl.
\bibitem{C92} F.Coester, Prog. Part. Nucl. Phys. {\bf 29} (1992) 1. 
\bibitem{SS} B. Silvestre-Brac and C. Semay, internal report, ISN
Grenoble, preprint ISN 93-69 (1993); C. Semay and B. Silvestre-Brac, Z.
Phys. C {\bf 61} (1994) 271; B. Silvestre-Brac, Few-Body Systems {\bf
20} (1996) 1.
\bibitem{SS98} C. Semay and B. Silvestre-Brac, Nucl. Phys. {\bf A 618}
(1997) 455.
\bibitem{BKSV94} B.Block, L.Koyrach, M.A.Shifman, and A.I.Vainstein, Phys. Rev. 
{\bf D49} (1994) 3356 [(E) {\bf D50} (1994) 3572].
\bibitem{NIR89} J.Nir, Phys. Lett. {\bf B221} (1989) 184.
\bibitem{JP} C.H.Jin and E.A.Paschos, Phys, Lett. {\bf B329} (1994)
364; C.H.Jin and E.A.Paschos, In {\it Proceedings of the
International Symposium on Heavy Flavor and Electroweak Theory},
Beijing, China,
1995, edited by C.H.Chang and C.S.Huang (World Scientific, Singapore,
1996), p.132.
\bibitem{BB93} G.Buchalla and A.J.Buras, Nucl. Phys. {\bf B400} (1993)
225.
\bibitem {BSW85} M.Bauer, B.Stech, and M.Wirbel, Z.Phys.
{\bf C29} (1985) 637, {\it ibid.} {\bf C34} (1987) 103.
\bibitem{NS97} M.Neubert, B.Stech, to appear in the Second 
Edition of {\it Heavy Flavours}, edited by A.J.Buras and 
M.Lindiner (World Scientific, Singapore), CERN--TH/97--99, 
hep--ph/9705292.
\bibitem{PDG98} Particle Data Group, C.Caso {\it et al},
Eur. Phys. J. {\bf C3} (1998) 1.
\bibitem{IS95} D.Scora, N.Isgur, Phys. Rev. {\bf D52} (1995) 2783.
\bibitem{S87} B.Stech, Phys. Rev. {\bf D36} (1987) 975.
\bibitem{DKND97} N.B.Demchuk, P.Yu.Kulikov, I.M.Narodetskii,
P.O'Donnell, Phys.
Lett. {\bf B409} (1997) 435 and Phys. Atom. Nucl., {\bf 60} (1997) 1292.
\bibitem{VS87} M.Voloshin and M.Shifman, Sov. J. Nucl. Phys. {\bf
45} (1987) 292
\end{thebibliography}
\end{document}